\documentclass[onecolumn,pra,preprint]{revtex4}%
\usepackage{amsfonts, amssymb,amsmath} % Include mathematical symbols
\usepackage{graphicx}% Include figure files
\usepackage{dcolumn}% Align table columns on decimal point
\usepackage{bm}% bold math

\newcommand{\be}{\begin{equation}}
\newcommand{\ee}{\end{equation}}
\newcommand{\bea}{\begin{eqnarray}}
\newcommand{\eea}{\end{eqnarray}}

\newcommand{\ti}{\i}
\newcommand{\tu}{\"u}
\newcommand{\tc}{\c c}

\newcommand{\tg}{\u g}

\newcommand{\tro}{\"o}

\newcommand{\tU}{\"U}
\newcommand{\tI}{\.I}
\newcommand{\Sch}{Schr\tro dinger equation~}

\newcommand{\ddf}{Dirac delta functions~}

\newcommand{\cf}{condensate fraction~}

\newcommand{\del}{$\delta$~}

\begin{document}

%\preprint{APS/123-QED}

\title{Bose-Einstein condensate in a harmonic trap with an eccentric dimple potential}

\author{Haydar Uncu$^1$, Devrim Tarhan$^2$, Ersan Demiralp$^{3,4}$,
\"{O}zg\"{u}r E. M\"{u}stecapl{\i}o\~{g}lu$^5$}
\affiliation{$^{1}$Department of Physics, Adnan Menderes
University, Kepezli Mevkii, 34342, Ayd\ti n, Turkey}
\affiliation{$^2$Department of Physics, Harran University,
Osmanbey Yerle\c{s}kesi ,63300, \c{S}anl\i urfa, Turkey}
\affiliation{$^{3}$Department of Physics, Bo\tg azi\tc i
University, Bebek, 34342, \tI stanbul, Turkey}
\affiliation{$^{4}$Bo\tg azi\tc i University-T\tU B\tI TAK Feza
G\tu rsey Institute Kandilli, 81220, \tI stanbul, Turkey}
\affiliation{$^5$Department of Physics, Ko\c{c} University
Rumelifeneri yolu, Sar\ti yer, 34450, \tI stanbul, Turkey}

\email{dtarhan@gmail.com}
\date{\today}
%-------------------------------------------------------------------------------
\begin{abstract}
We investigate  Bose-Einstein condensation of noninteracting gases
in a harmonic trap with an off-center dimple potential.  We
specifically consider the case of a tight and deep dimple
potential which is modelled by a point interaction. This  point
interaction is represented by a Dirac delta function.   The atomic
density, chemical potential, critical temperature and condensate
fraction, the role of the relative depth and the position of the
dimple potential are analyzed by performing numerical
calculations.

{\bf Topic:} Physics of Cold Trapped Atoms

{\bf Report number:} 6.13.1
\end{abstract}
%---------------------------------------------------------------------
%------------------------------------------------------------------------
\maketitle
%\narrowtext
%==============================================================================

\section{INTRODUCTION}

The phase space density of a Bose-Einstein condensate (BEC) can be
increased by modification of the shape of the potential
\cite{pinkse}. ``Dimple"-type potentials are the most commonly
used potentials for this purpose \cite{kurn,weber,ma}.  A small
dimple potential at the equilibrium point of the harmonic trapping
potential is used to enhance phase-space density  by an arbitrary
factor \cite{kurn}. A tight dimple potential is used for a recent
demonstration of caesium BEC  a \cite{weber}. Quite recently, such
potentials were proposed for efficient loading and fast
evaporative cooling to produce large BECs \cite{comparat}.
Attractive applications, such as controlling interaction between
dark solitons and sound \cite{parker}, introducing defects such as
atomic quantum dots in optical lattices \cite{jaksch}, or quantum
tweezers for atoms \cite{diener} are offered by using tight dimple
potentials for (quasi) one-dimensional BECs . Such systems can
also be used for spatially selective loading of optical lattices
\cite{griffin}. In combination with the condensates on atom chips,
tight and deep dimple potentials can lead to rich novel dynamics
for potential applications in atom lasers, atom interferometers
and in quantum computations (see Ref. \cite{proukakis} and
references therein).

In this paper we continue the discussion of our recent paper
\cite{collective}. In that paper, we modelled the dimple type
potentials by Dirac \del functions and investigated the change of
chemical potential, critical temperature and condensate fraction
of a harmonic trap with respect to the various strengths of Dirac
$\delta$ functions.  In this paper, we investigate the behavior of
the same physical quantities for a $\delta$ function which can be
located at different positions than the center of a harmonic trap.
We find that while a centrally positioned dimple potential is most
effective in large condensate formation at enhanced temperatures,
there is a critical location for which the condensate fraction and
the critical temperature can also be enhanced relatively. This
might be useful in spatial fragmentation of atomic condensates.

The paper is organized as follows. In Sect. II,  we review shortly
the analytical solutions of the \Sch for a harmonic potential with
a finite number of Dirac $\delta$-decorated harmonic potential and
give the eigenvalue equation of the harmonic potential with a
Dirac \del function. In Sect. III, determining the eigenvalues
numerically, we show the effect of the dimple potential on the
condensate fraction and the transition temperature and investigate
the change of this values with respect to the position of the
Dirac \del function. Finally, we  conclude in Sect. IV.

%-------------------------------------------------------------------------------

\section{Harmonic potential Decorated with \ddf}

We begin our discussion by reviewing the one dimensional harmonic
potential decorated with the Dirac \del functions
\cite{collective}-\cite{demiralp3} for the seek of completeness.
The potential for  is given as:
\be V(x)=\frac{1}{2} m \omega^2 x^2-\frac{\hbar^2}{2m} \sum_i^P
\sigma_i \delta(x-x_i), \label{potential} \ee
where $\omega$ is the frequency of the harmonic trap, $P$ is a
finite integer and $\sigma_i$'s are the strengths (depths) of the
dimple potentials located at $x_i$'s with $x_1<x_2<...<x_P$ with
$x_i \in (- \infty,\infty)$. The factor $\hbar^2/2m$ is used for
calculational convenience. Negative $\sigma_i$ value represents
repulsive interaction while positive $\sigma_i$ value represents
attractive interaction. The time-independent \Sch equation for
this potential is written as:
\be -\frac{\hbar^2}{2m} \frac{d^2 \Psi(x)}{dx^2} + V(x) \Psi(x)= E
\Psi(x). \label{Scheq} \ee
By defining $E=(\xi+\frac{1}{2}) \hbar \omega$, with $\xi$ a real
number, and introducing dimensionless quantities $z=x/x_0$, and
$z_i=x_i/x_0$ with $x_0=\sqrt{\hbar/2m \omega}$, the natural
length scale of the harmonic trap, we can re-express Eq.
(\ref{Scheq}) as
\be
 \frac{d^2 \Psi(z)}{dz^2}+ \left [ \xi + \frac{1}{2}-\frac{z^2}{4}
 +  \sum_i^P \Lambda_i \delta(z-z_i)\right ] \Psi(z) =0,
 \label{parcyldiffeq}
\ee
where $\Lambda_i=x_0\sigma_i$.  By using transfer matrix approach
\cite{collective,demiralp3, demiralp}, we get the following
eigenvalue equation:
\be 1-\frac{ \Lambda_1 D_{\xi}(z_1) \, D_{\xi}(-z_1)}{W}=0
\label{eigenvaleq}
\ee
for $\xi$ and using $E=(\xi+\frac{1}{2}) \hbar \omega$. Here,
 $D_{\xi}(z)$ and $ D_{\xi}(-z) $ are parabolic cylinder functions
 and
$z_1=x_1/x_0$. The  Wronskian $W$ of $ D_{\xi}(z)$ and $
D_{\xi}(-z)$ is
\be
W=W[D_{\xi}(z),D_{\xi}(-z)]=\frac{2^{(\xi+3/2)}
\pi}{\Gamma\left(-\frac{\xi}{2} \right) \, \Gamma\left(\frac{1-
\xi}{2} \right) }
\ee
For $z_1=0$, these results reduce to the results in Ref.
\cite{collective,demiralp3, avakian}.

%%%%%%%%%%%%%%%%%%%%%%%%%%%%%%%%%%%%%%%%%%%%%%%%%%%%%%%%%%%%%%%%%%%%%%%%%%%%%%%
%%%%%%%%%%%%%%%%%%%%%%%%%%%%%%%%%%%%%%%%%%%%%%%%%%%%%%%%%%

\section{BEC In a One-Dimensional Harmonic Potential with a Dirac \del Function}

In this section we calculate the condensate fraction, chemical
potential, critical temperature and density profile for different
depths, sizes and positions of a dimple potential modelled by a
Dirac \del function. In order to describe the depth and size of a
dimple potential in a systematic way we define a dimensionless
variable in terms of the strength of the Dirac \del functions as:
\be \Lambda= \sigma \sqrt{ \frac{\hbar}{2m \, \omega}}\, .
\label{lambda}
\ee
We will present our results with respect to $\Lambda$ and $z_1$
defined in the previous section.

In ref. \cite{collective}, we have estimated $\sigma$ values
approximately according the parameters in refs. \cite{hau} and
\cite{weber}. We find that, if $ 10^8 \, 1/ \mathrm{m} \leq \sigma
\leq 10^{10} \, 1/ \mathrm{m} $ then $ 320 \leq \Lambda \leq
32000$ for the experimental parameters m$=23$ amu
($^{23}\mathrm{Na}$), $\omega= 2 \pi\times 21$ Hz \cite{hau} and
for the experimental parameters m$=133$ amu ($^{133}\mathrm{Cs}$),
$\omega= 2\pi \times 14$ Hz \cite{weber}. In ref.
\cite{collective}, we  have shown that, even for small $\Lambda$
values, \cf and critical temperature change considerably. How
sensitive such a change would occur depending on the location of
the dimple trap was a question left unanswered in
Ref.\cite{collective}.

We begin our discussion by investigating the change of the
critical temperature with respect to the position of Dirac
$\delta$ function.
\begin{figure}
{\vspace{0.5cm}}
\includegraphics[width=3.5 in]{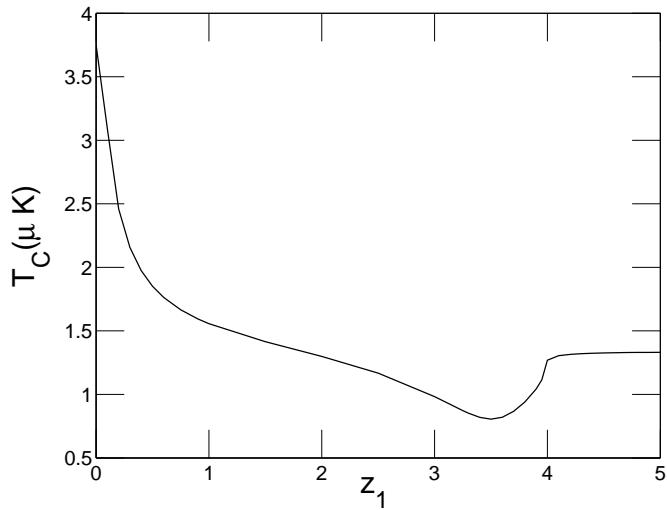}
\caption{ The critical temperature $T_c$ for $N=10^4$. $\Lambda $
is a dimensionless variable defined in Eq. (\ref{lambda}). Here we
use m $=23$ amu ($^{23}\mathrm{Na}$) and $\omega= 2 \pi \times 21$
Hz \cite{hau}.} \label{critemp}
\end{figure}
The critical temperature ($T_c$) is obtained by taking the
chemical potential equal to the ground state energy
($\mu=E_g=E_0$) and by solving
\be N \approx \sum_{i=1}^{\infty} \frac{1}{e^{\beta_c
\varepsilon_i}-1} \;\; , \label{Tcritical} \ee
for $\beta_c$ where $\beta_c=1/(k_B T_c)$. For finite $N$ value,
we define $T_c^0$ as the solution of Eq. (\ref{Tcritical}) for
$\Lambda=0$ (only the harmonic trap).

In Eq. (\ref{Tcritical}),  $ \varepsilon_i$'s are the eigenvalues
for the harmonic potential decorated with a single eccentric
dimple potential at $z_1$. The energies of the decorated states
are found by solving Eq. (\ref{eigenvaleq}) numerically. Then,
these values are substituted into the Eq. (\ref{Tcritical}); and
finally this equation is solved numerically to find $T_c$. We
obtain $T_c$ for different $z_1$ values and $ \Lambda=32 $ and
show our results in Fig. (\ref{critemp}). Since harmonic potential
is symmetric negative $z_1$ values will give a the same values for
critical temperature with positive ones. As $z_1$ increases, the
energy of the ground state increases so that the critical
temperature decreases as $z_1$ gets larger. On the other hand, as
the dimple trap becomes farther to the center of the harmonic
trap, the critical temperature cease to decrease and starts rising
again as seen in Fig. (\ref{critemp}) around $z_1=3.5 $. Finally,
at very large separations between the dimple trap and the harmonic
trap center, the critical temperature no longer changes with the
location of the dimple trap and saturates at the value
corresponding to the of the critical temperature for the single
harmonic trap per se. The increase of the critical temperature
around $z_1=3.5 $ can be explained as follows: As $z_1$ increases
it becomes closer to the node of the first exited state wave
function of the harmonic potential. At that value the change of
the energy eigenvalue of the first excited state vanishes and the
difference between the first excited state and ground state
eigenvalues increase. Thus, the particles favor the ground state
which increases the critical temperature. In Fig. (\ref{critemp})
we take $N=10^4$ and use typical experimental parameters m$=23$
amu ($^{23}\mathrm{Na}$) and $\omega= 2\pi \times 21$ Hz
\cite{hau}.

\begin{figure}
\centering{\vspace{0.5cm}}
\includegraphics[width=3.5 in]{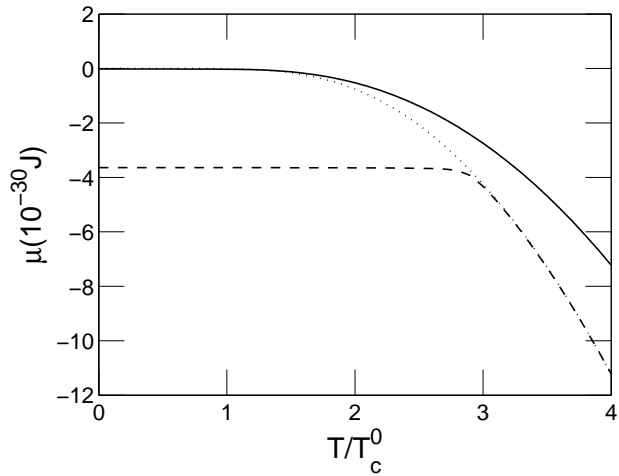}
\caption{The chemical potential  vs temperature $T/T_c^0$ for
$N=10^4$ . The solid line shows $\mu$ only  for the harmonic trap.
The dotted line shows $\mu$ for $z_1=1$ and $\Lambda=32$. The
dashed line shows $\mu$ for $z_1=0$ and $\Lambda=32$ . The other
parameters are the same as Fig. \ref{critemp}.} \label{muvstgraph}
\end{figure}

For a gas of N identical bosons, the chemical potential $\mu$ is
obtained by solving
\be N=\sum_{i=0}^{\infty} \frac{1}{e^{\beta (\varepsilon_i -
\mu)}-1}= N_0 + \sum_{i=1}^{\infty} \frac{1}{e^{\beta
(\varepsilon_i - \mu)}-1}, \label{chem pot} \ee
at constant temperature and for given N, where $\varepsilon_i$ is
the energy of state $i$. We present the change of $\mu $ as a
function of $T/T_c^0$ in Fig.(\ref{muvstgraph}) for $ N=10^4 $; $
\Lambda = 0$; $ \Lambda = 32 $, $z_1=0$ and  $ \Lambda = 32 $
$z_1=1$.

\begin{figure}
\centering{\vspace{0.5cm}}
\includegraphics[width=3.5in]{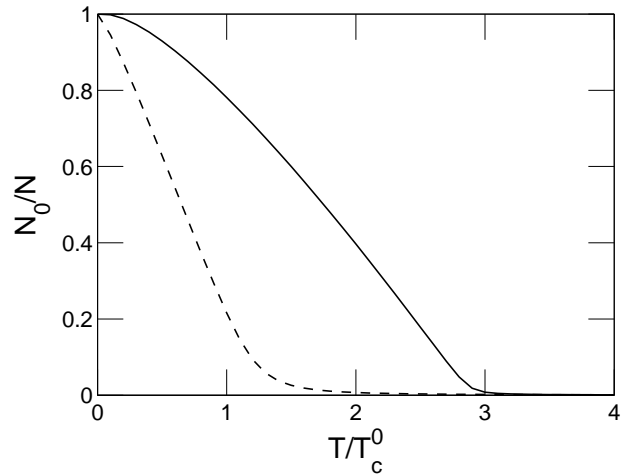}
\caption{$N_0/N$ vs $T/T_c^0$ for $N=10^4$ and $\Lambda=32$. The
solid line for $z_1=0$, the dashed line for $z_1=1$. The other
parameters are the same as Fig. \ref{critemp}.} \label{condfrac}
\end{figure}

By inserting $\mu $ values into the equation
\be N_0= \frac{1}{e^{\beta (\varepsilon_0 - \mu)}-1},\ee
we find the average number of particle in the ground state.
$N_0/N$ versus $T/T_c^0$ for $N=10^4$ $\Lambda= 32 $  are shown in
Fig. (\ref{condfrac}). In this figure, the solid line shows the
condensate fraction for $z_1=0$ and the dashed line shows the
condensate fraction for $z_1=1$.  In Ref. \cite{ketterle},
Ketterle et. al mentions that the phase transitions due to
discontinuity in an observable macro parameter occurs only in
thermodynamic limit, where $N \rightarrow \infty$. However, we
make our calculations for a realistic system with a finite number
of particles in a confining potential. Thus, $N_0/N$ is a finite
non-zero quantity for $T<T_c$ without having any discontinuity at
$T=T_c$.

\begin{figure}
\centering{\vspace{0.5cm}}
\includegraphics[width=3.5in]{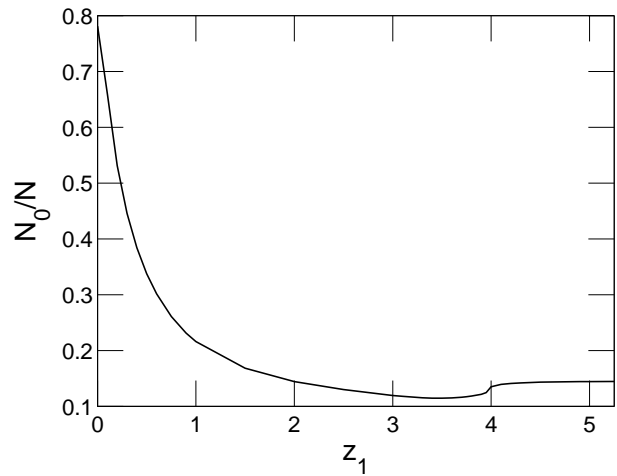}
\caption{$N_0/N$ vs $T/T_c^0$ for $N=10^4$ and $\Lambda=32$. The
other parameters are the same as Fig. \ref{critemp}.}
\label{condfrac2}
\end{figure}

We also investigate the behavior of the condensate fraction as a
function of the position of Dirac \del for $\Lambda=32$, $T=T_c^0$
and present the results in Fig. (\ref{condfrac2}).

\begin{figure}
\centering{\vspace{0.5cm}}
\includegraphics[width=3.5in]{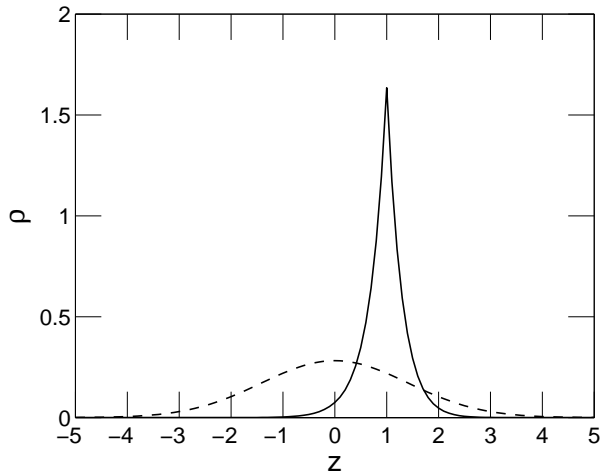}
\caption{Comparison of  density profiles of a BEC in a harmonic
trap with a BEC in a harmonic trap decorated with a \del function
($\Lambda=3.6, \; z_1=1$). The solid curve is the density profile
of the BEC in decorated potential. The dashed curve is the density
profile of the 1D harmonic trap ($\Lambda=0$). The parameter z is
dimensionless length defined after Eq. (\ref{Scheq}). The other
parameters are the same as Fig. \ref{critemp}.} \label{Denprof}
\end{figure}

Finally, we compare density profiles of condensates for a harmonic
trap and a harmonic trap decorated with a delta function ($\Lambda
=3.2$ and $z_1=1$)  in Fig. (\ref{Denprof}). Since the ground
state wave functions can be calculated analytically for both
cases, we find the density profiles by taking the absolute square
of the ground state wave functions. Comparing the graphics of
density profiles, we see that an offcenter dimple potential
maintain a higher density at the position of the Dirac \del
function which may be utilized for the fragmentation of a BEC.

\section{Conclusion}

We have investigated the effect of the location of the tight
dimple potential on the results reported recently in our paper
\cite{collective}. We model the tight dimple potential with the
Dirac \del function. This allows for analytical expressions for
the eigenfunctions of the system and simple eigenvalue equations
greatly simplifies subsequent numerical treatment. We have
calculated the critical temperature, chemical potential,
condensate fraction and presented the effects of the location of
the dimple potential. We find that the dimple type potentials are
most effective when they are applied to the center. While it is
also advantageous to place the dimple potential at the nodes of
the excited state, where our results revealed a relative
enhancement of the critical temperature and the condensate
fraction. Determining the density profiles of the BECs in the
harmonic trap and in the decorated trap with the Dirac \del
function at this critical position, we argued that eccentric
dimple trap at such a critical location can be used for spatial
fragmentation of large, enhanced BECs.

The presented results are obtained for the case of noninteracting
and one dimensional condensates for simplicity. In such a case,
stability of the condensate may become questionable and should be
addressed separately in detail. \cite{yukalov}. The treatment
should be extended for the case of interacting condensates in
larger (or quasi) dimensional traps in order to make the results
more relevant to experimental investigations.

%%%%%%%%%%%%%%%%%%%%%%%%%%%%%%%%%%%%%%%%%%%%%%%%%%%%%%%%%%%%%%%%%%%%
%%%%%%%%%%%%%%%%%%%%%%%%%%%%%%%%%%%%%%%%%%%%%%%%%%%%%%%%%%%%%%%%%%%%
%%%%%%%%%%%%%%%%%%%%%%%%%%%%%%%%%%%%%%%%%%%%%%%%%%%%%%%%%%%%%%%%%%%%

\acknowledgments

H.U. gratefully acknowledges illuminating comments and discussions
by V.I. Yukalov. O.E.M. acknowledges the support from a
T\"UBA/GEB\.{I}P grant. E.D. is supported by Turkish Academy of
Sciences, in the framework of the Young Scientist Program (ED-
T\"UBA- GEBIP-2001-1-4).

%%%%%%%%%%%%%%%%%%%%%%%%%%%%%%%%%%%

\newpage

{\bf Figure Captions}

{\bf Fig. 1} The critical temperature $T_c$ for $N=10^4$. $\Lambda
$ is a dimensionless variable defined in Eq. (\ref{lambda}). Here
we use m $=23$ amu ($^{23}\mathrm{Na}$) and $\omega= 2 \pi \times
21$ Hz \cite{hau}.

{\bf Fig. 2} The chemical potential  vs temperature $T/T_c^0$ for
$N=10^4$ . The solid line shows $\mu$ only  for the harmonic trap.
The dotted line shows $\mu$ for $z_1=1$ and $\Lambda=32$. The
dashed line shows $\mu$ for $z_1=0$ and $\Lambda=32$ . The other
parameters are the same as Fig. \ref{critemp}.

{\bf Fig. 3} $N_0/N$ vs $T/T_c^0$ for $N=10^4$ and $\Lambda=32$.
The solid line for $z_1=0$, the dashed line for $z_1=1$. The other
parameters are the same as Fig. \ref{critemp}.

{\bf Fig. 4} $N_0/N$ vs $T/T_c^0$ for $N=10^4$ and $\Lambda=32$.
The other parameters are the same as Fig. \ref{critemp}.

{\bf Fig. 5} Comparison of  density profiles of a BEC in a
harmonic trap with a BEC in a harmonic trap decorated with a \del
function ($\Lambda=3.6, \; z_1=1$). The solid curve is the density
profile of the BEC in decorated potential. The dashed curve is the
density profile of the 1D harmonic trap ($\Lambda=0$). The
parameter z is dimensionless length defined after Eq.
(\ref{Scheq}). The other parameters are the same as Fig.
\ref{critemp}.

\newpage
{\bf Corresponding author:} Devrim Tarhan\\
Office Phone: +90-414-344 0020/1348\\
Fax: +90-414-344 0051\\
E-Mail: dtarhan@gmail.com

\end{document}